\documentclass[letterpaper,twocolumn,10pt]{article}
 \pdfoutput=1
 \usepackage{usenix2025_SOUPS}
 \usepackage{epsfig,endnotes,listings}
 \usepackage{pifont}
 \usepackage{MnSymbol}
 \usepackage{graphicx}
 \usepackage{caption}
 \usepackage{booktabs}
 \usepackage{tabularx}
 \usepackage{xcolor}
 \usepackage{placeins}
 \usepackage{listings}
 \usepackage{censor}
 \newcommand{\redact}[1]{\censor{#1}}
\renewcommand{\redact}[1]{#1}

 \definecolor{codegreen}{rgb}{0,0.6,0}
 \definecolor{codegray}{rgb}{0.5,0.5,0.5}
 \definecolor{codepurple}{rgb}{0.58,0,0.82}
 \definecolor{backcolour}{rgb}{0.95,0.95,0.92}

 \lstdefinestyle{codeStyle}{
     commentstyle=\itshape,
     keywordstyle=\bfseries,
     numberstyle=\tiny\sffamily,
     stringstyle=\ttfamily,
     basicstyle=\ttfamily\footnotesize,
     breakatwhitespace=false,         
     breaklines=true,                 
     keepspaces=true,                 
     numbers=left,       
     numbersep=5pt,                  
     showspaces=false,                
     showstringspaces=false,
     showtabs=false,                  
     tabsize=2,
 }

 \newcommand{\credit}[1]{{\small\sffamily---#1}}

 \newcommand{\tablestyle}[0]{\sffamily\footnotesize\centering}

 \newcommand{\code}[3]{
     \lstinputlisting[
         caption=#1,
         label={lst:#1},
         language=#2,
         style=codeStyle,
     ]{#3}
 }

 \newcommand{\xmark}{\ding{55}}
 \newcommand{\cmark}{\ding{51}}

\begin{document}
 \title{\Large\bf
   Playing in the Sandbox:
   A Study on the Usability of Seccomp
 }

\def\plainauthor{Maysara Alhindi and Joseph Hallett}
\author{{\rm Maysara Alhindi} \\ University of Bristol \and {\rm Joseph Hallett}\\ University of Bristol}

 \maketitle
 \thecopyright

 \begin{abstract}
    Sandboxing restricts what applications do, and prevents exploited processes being abused; yet relatively few applications get sandboxed: why?  We report a usability trial with 7 experienced Seccomp developers exploring how they approached sandboxing an application and the difficulties they faced. The developers each approached sandboxing the application differently and each came to different solutions. We highlight many challenges of using Seccomp, the sandboxing designs by the participants, and what developers think would make it easier for them to sandbox applications effectively.
 \end{abstract}

 \section{Introduction}

 A backdoor was discovered in one of the most popular compression libraries, \texttt{xz}~\cite{lins2024critical}. Whilst several security techniques (like sandboxing) could have mitigated the attack and prevented exploitation, these techniques were not always used. In the case of the \texttt{xz} backdoor, sandboxing would limit the attack surface as the library would be able access the functionality it needs, and nothing more. 

 Despite the security benefits of sandboxing, it is not always incorporated into applications, with different OSs sandboxing different things.  OpenBSD sandboxes the \texttt{xz} application by default using its \emph{Pledge} sandboxing mechanism, and whilst the \texttt{xz} exploit did not target OpenBSD, the same techniques would not have made the backdoor useful in the same way as it was with Linux. Despite Linux having a more capable sandboxing mechanisms than OpenBSD (specifically, Seccomp), \texttt{xz} is not as restricted on Linux by default.  One study on the number of sandboxed packages available in different OSs found that only 802 packages for the popular Debian and Fedora Linux OSs directly used sandboxing in their code out of the tens of thousands of packages available~\cite{alhindi2024sandboxing}.

 One reason for the relatively low sandbox adoption is that Seccomp (the Linux sandboxing framework) is \emph{too} powerful---it provides a general programming language for sandboxing with an assembly-esque syntax that can be used by developers in many different ways and can be difficult to understand and employ correctly---even for developers who have used Seccomp before.
 Developers have to make design decisions when using Seccomp: they must decide what they want to sandbox in a program, with what permissions, and why. By examining how existing developers use Seccomp, we can better understand how they sandbox their code and help make it easier for others to add sandboxes to their own code.

 This paper reports a usability study with 7 developers who had used Seccomp before, that were tasked with sandboxing a program. We observed them working and asked questions as they worked, alongside an interview and design exercise.
 
 \vspace\baselineskip\noindent
 With developers with prior exposure to Seccomp, we ask the following research questions:
 \begin{itemize}
 \item Do existing developers understand sandboxing and employ Seccomp effectively?
 \item What aspects of Seccomp do developers struggle with when sandboxing their applications?
 \item What do developers want from a sandbox?
 \end{itemize}

 We find that the developers took different approaches to sandboxing their programs, both in terms of the implementations and the way they arrived at them. Developers struggled with aspects of Seccomp, such as working out what syscalls were being used where.  To the best of our knowledge, this represents the first usability study of the Seccomp sandbox and the only usability study examining how developers utilize sandboxing mechanisms to secure their code.

 \section{Background and Related Work}

 \subsection{Developer-centric sandboxing}

 There are many sandboxing and access control solutions that help in achieving the \emph{principle of the least privilege}~\cite{Saltzer_Schroeder_1975}. Some, such as SELinux and AppArmor, are configured with a security policy and restrict the resources applications can access~\cite{schreuders2013state}. These solutions are mainly designed to be used by system administrators as they require configuring external security policies without altering an application's code.

 Other solutions are designed for developers. They offer developers APIs that can be integrated into an application's codebase, allowing for the limitation of its access to specific system resources such as syscalls and the filesystem. These mechanisms let developers sandbox specific code segments and drop privileges programmatically. Unlike app-oriented sandboxes, such as AppArmor, the sandboxing policy exists within the code. Different operating systems have different sandboxing APIs, Linux has Seccomp~\cite{seccomp}, OpenBSD has Pledge~\cite{pledge}, while FreeBSD provides Capsicum~\cite{capsicum-cap-for-unix}.

 \code{Seccomp example}{C}{code/seccomp_example.c}

 Seccomp (short for Secure Computing) is a Linux sandboxing mechanism used by many applications (including Chrome, Firefox and OpenSSH~\cite{alhindi2024sandboxing}) that lets developers restrict what syscalls their programs can make. Developers can use Seccomp to allow, deny, or trap selected syscalls and check syscall arguments.  For example, developers can allow their programs to write to only the standard output by restricting the arguments of the \texttt{write} syscall (Listing~\ref{lst:Seccomp example}). Developers can either write raw BPF filters or use wrappers like \texttt{libseccomp} to implement their sandbox.

 \subsection{Prior work on sandboxing}

 Sandboxing research is broad, and reviews generally cover broader techniques rather than specific sandboxing implementations.
 Maass~et~al{.} conducted an interdisciplinary systematic literature review of general sandboxing techniques, highlighting numerous research gaps, particularly in validation techniques and the usability aspects~\cite{maass2016systematic}. Schreuders~et~al{.} categorized sandboxes into rule-based and isolation-based types, comparing their functionalities and limitations~\cite{schreuders2013state}. These reviews highlighted the importance of the usability of sandboxing and stressed that it is under-researched~\cite{Maass-thesis}.

 Research into the APIs that sandboxes provide is less varied.
 Research by Anderson compares Seccomp, Pledge, and Capsicum, focusing on their technical differences and implementation~\cite{anderson2017comparison}. Alhindi~et~al{.} investigated the adoption of sandboxing mechanisms in different OSs, comparing how these sandboxes worked across different operating systems~\cite{alhindi2024sandboxing}. They found that fewer than 1\% of packages directly use these mechanisms and that fine-grained mechanisms like Seccomp are sometimes used to mimic simpler models like Pledge, and highlighted a gap in understanding developers' needs and the usability of these mechanisms.

 \subsection{Usability of Security APIs}

We found minimal research looking at the usability of sandboxing APIs---the closest research to this was done by Schreuders~et~al{.} where they examined the usability of access control mechanisms such as AppArmor and SELinux~\cite{schreuders2012towards}. Their research highlighted negative themes that affect the usability of these systems such as the complexity of the sandboxing policy, and proposed suggestions to improve them, including better descriptions and documentation, and better integration with automation tools and the overall operating system. Other research investigated how sandboxed apps could interfere with the needs of desktop users by conducting interviews with 13 expert Linux users~\cite{dodier2017no}. The research shows that users could abandon apps if developers removed features for security reasons; such as implementing sandboxing. They argue that sandboxes sometimes require developers to remove some features that are important for the functionality of applications in order to sandbox them (such as plugins) and that this could be why developers do not adopt sandboxes, as the ones who choose to do so, might have to drop features, and thus lose users. 

Neglecting usability has consequences~\cite{acar2017comparing, yee2004aligning}. Many security APIs suffer from usability issues, making them difficult for developers to use effectively~\cite{acar2017comparing} with prior work primarily focusing on cryptography APIs~\cite{green2016developers}.  Green and Smith developed principles for usable cryptography APIs~\cite{green2016developers}. Patnaik~et~al{.} examined over 2,400 StackOverFlow questions related to cryptography APIs to understand the issues developers face when using these APIs and how Green and Smith’s usability principles relate to these issues~\cite{patnaik2019usability}. Research by Acar~et~al{.} assessed the usability of different cryptographic libraries by conducting a controlled study with 256 Python developers aiming to understand how the usability of these libraries affects the security of the implemented code. Their results found that many participants believed their solutions were secure while it was not, and that participants struggled with simple tasks. They highlighted the importance of having good documentation and code examples~\cite{acar2017comparing}.

 While this research touches on many aspects of sandboxes and usability, there is no research that examines the usability of sandboxing APIs and attempts to understand developers' rationale when using them. 
Other studies have examined the usability of other security tools including firewalls~\cite{voronkov2020measuring}, the Certbot certificate tool~\cite{tiefenau2019certbot}, and HTTPS~\cite{krombholz2017https} and fuzzers~\cite{ploger2023afl}.  This paper adds to a growing body of work examining how developers interact with security focussed code, and trying to improve usability for developers.

 \section{Methodology}

 To examine Seccomp's usability, we recruited 7 developers, all with prior experience of Seccomp (Table~\ref{tab:Seccomp_participants_background}). We recruited participants through kernel security mailing lists and mailing lists of software using Seccomp. We also posted the study details on freelancing platforms such as Upwork and Freelancer. Additionally, we approached developers on related Linux security forums, contacted developers from our professional network, and reached out to researchers who have done research on Seccomp. We compensated participants with approximately \$250 in their local currency for 1.5 hours of their time to ensure that it is above the average hourly rate, a reasonable salary for a job, and attractive for skilled developers.  Our rejection criteria included any developers who reported not having written code with Seccomp before:
 we asked developers to tell us about their Seccomp and C programming experience, and we excluded 4 developers who did not have experience with sandboxing programs using Seccomp but were willing to participate.  Table~\ref{tab:Seccomp_participants_background} lists our participants' experience with Seccomp.  All 7 participants were male, and were based all over the world. Further demographics were not collected as it was discouraged by our ethics agreement.

 \begin{table}
   \tablestyle
   \centering
   \caption{Participants' experience with Seccomp.}
   \begin{tabularx}{\linewidth}{l X}
           \toprule
           P1 & Exploit programs that are poorly sandboxed \\
           P2 & Teach Seccomp in a university lab\\
           P3 & Sandbox an open-source networking package\\
           P4 & Sandbox client-server software\\
           P5 & Sandbox automotive embedded-systems software\\
           P6 & Experimental usage of Seccomp and for blogging\\
           P7 & Create an open-source container software\\
           \bottomrule 
       \end{tabularx}
     \label{tab:Seccomp_participants_background}
 \end{table}

 The sessions were conducted online via Zoom and took 1.5 hours. Prior to the sessions, participants were given access to a remote Linux virtual machine either using SSH or VNC. We asked participants if they would like any particular software or code editors installed on the machine so they could feel comfortable and focus on solving the task. The VM had all the needed libraries and compilers to compile C programs with Seccomp. Audio of the participants from the study was recorded and transcribed. We asked participants to \emph{think aloud}~\cite{ibm-rc-9265} and express any thoughts or struggles, we also took note of any interesting behavior and references to documentation or resources. We asked participants to share their screens with us while they were solving the task, but the screen was not recorded. We gave participants the space to implement their sandbox, only giving when they were stuck for a while: which P4 and P5 needed.  Our full study protocol is in Appendix~\ref{appendix:protocol}

 \subsection{Study sessions and tasks}

 We used semi-structured interviews because they offer flexibility and enable us to gather detailed data. While we followed a standard protocol for all interviews, we also asked additional clarifying questions when a participant exhibited interesting behavior to better understand their perspectives and reasoning. To refine our study procedure, we conducted 3 pilot sessions and modified the interview script based on feedback from these trial interviews.  The data from these sessions is not included in our analysis. The study was comprised of 3 tasks. 

 \subsubsection{Sandboxing a watchdog script}

 In the first task, we asked developers to act as an administrator and sandbox a part of their critical infrastructure using Seccomp, focusing on security and stability. The program provided is 115 lines of C and simulates a watchdog service, it handles signals, checks if a port is open on a remote server, stores the results in a log file, and rotates the log file every few seconds by checking its size and renaming it at regular intervals. The code of the watchdog script is included in Appendix~\ref{appendix:code}. Developers could modify the C code and implement the sandbox however they liked. They were given 1 hour for the task and could access any resources and documentation they liked. 

 We noted the developers' approaches to the task, any errors they made, and the documentation referenced. We noted signs of frustration, motives, and their train of thought. During this phase, we asked developers about their decisions regarding:
 \begin{enumerate}
     \item The use of a deny-list or allow-list approach to Seccomp
     \item Examining syscall's arguments and limiting them
     \item Restructuring the code to implement a stricter sandbox
     \item Where in the program they decided to install Seccomp filters
 \end{enumerate}

 \subsubsection{Filling a System-Usability-Scale form}

 The second task involved filling a System-Usability-Scale (SUS) form~\cite{lewis2018system}. The form questions were modified to be about Seccomp. We asked developers to briefly expand on their answers to each question. SUS has been used in many usability studies before, and used in security related usability studies~\cite{dewitt2006aligning}. It features 10 questions rated on a Likert scale touching on aspects of the studied system (such as its complexity and consistency, and the overall users' experience when using the system). We used SUS because it covers many usability aspects of the studied system and it is short and easy to fill, which avoid fatigue or frustration leading to inaccuracy.

 \subsubsection{Participatory Design}

 In the final task, we shared a whiteboard with participants and asked them to tell us about their ideal sandboxing model. We asked participants to design a sandbox that can be used within a codebase, and to focus more on the overall design and interface of it rather than its implementation. Participants could describe their thoughts verbally or by drawing on the shared whiteboard. Participatory design~\cite{spinuzzi2005methodology} is a tool to involve participants in the design process and get their ideas and suggestions and has been used to design static analysis tools~\cite{johnson2013don}. As our participants have sandboxing experience we believe they are best placed to identify the designs that would work for them.

 \subsection{Analysis}

 Sessions were transcribed and anonymized manually to help the researcher familiarize themselves with them, before being qualitatively coded to extract themes. We used open coding~\cite{strauss2004open} on the transcripts, assigning codes to key concepts and patterns that emerged from the data. As new codes emerged, we iterated through the previous sessions to ensure consistency and that all relevant patterns were captured, allowing themes to evolve throughout the analysis process.

 We based our analysis on ISO-9241 which defines usability as: \emph{``The extent to which a system, product or service can be used by specified users to achieve specified goals with effectiveness, efficiency, and satisfaction in a specified context of use''}~\cite{bevan2015iso}, and used these as apriori axial codes to structure our findings (Section~\ref{sec:results}).

 \subsection{Threats to Validity}

 There are several limitations to our study. Our sample size is tiny and does not generalize. Recruiting developers for usability studies is known to be hard~\cite{patnaik2022if} and recruiting developers with specialist experience is particularly challenging. The recruitment process took several months, we reached out to tens of developers, published the details of the study on many platforms and websites, and ensured that developers would be compensated well for their time, but very few developers matched our criteria of having experience with Seccomp.

 While we did not reach full theoretical saturation, the findings remain valuable---if not generalizable. We tried to extract as much data from each session as possible using methods that work well with small sample sizes~\cite{ibm-rc-9265}. The study sessions incorporate multiple data sources (Triangulation)~\cite{triangulation2014use}, such as the coding task, think-aloud protocol, and the participatory design exercise. Our analysis offers in-depth insights into the developers' thought processes~~\cite{ponterotto2006brief}, as we assess the effectiveness of the implemented sandbox and describe the reasoning behind developers' decisions, the struggles they faced while implementing it, and record developers' suggestions and designs. Despite the small sample size, we still obtained useful information about how developers approach sandboxing. While it may not generalize (and we do not claim that it does) qualitative approaches can still lead to new insights even with small sample sizes~\cite{virzi1992refining,nielsen2000you}.

 When designing the coding task, we tried to make it as close to a real-life situation as possible by tasking participants with a code that simulates a watchdog service, a type of service that is common. We also placed the participants in a security-critical scenario to motivate them to sandbox the program to their best capabilities, but even with that, participants' performance could still be somewhat different from what it could be in real life. Participants' performance could have been affected by the Hawthorne Effect, which occurs when participants realize that they are being watched~\cite{jones1992there}, or by the Social Desirability Bias~\cite{grimm2010social}, as participants could feel inclined to perform better to match researchers' expectations or desires. We told participants that the audio recording would be destroyed and that their sessions would be anonymized, which could help in reducing these effects.

 \section{Results}
 \label{sec:results}

 \subsection{Effectiveness}

 Do developers understand sandboxing and employ Seccomp effectively? 
 We observed the developers attempts to sandbox the watchdog service and recorded their design decisions. For a simple program like the watchdog script, we assumed that developers' implementations would not vary, as the program is short and simple, and does not contain any dynamic code or hidden logic. It was surprising to us that developers' implementations of Seccomp sandbox were \emph{completely} different. All bar one (P4), of our participants had a working solution. Table \ref{tab:Seccomp_implementation} summarises how each implementation differed.

 \begin{table*}
   \tablestyle
   \centering
   \newcommand{\allow}[0]{$\medsquare$}
   \newcommand{\deny}[0]{$\filledmedsquare$}
   \caption{Seccomp sandbox implementation details and approaches for each participant, including whether they applied their sandbox to the program's main loop, or the server's main function.}
     \begin{tabularx}{\linewidth}{lXXXXXXX}
           \toprule
           Criteria &  P1 & P2 & P3 & P4 & P5 & P6 & P7  \\
           \midrule
           Functional Sandbox?                       & \cmark  & \cmark & \cmark & \xmark   & \cmark & \cmark   & \cmark \\
           Allow/Deny (\allow/\deny) approach & \allow  & \allow & \allow & \deny    & \allow & \allow   & \allow  \\
           Syscalls filtered            & 9       & 11     & 22     & 1        & 26     & 11       & 27  \\
           Sandbox applied to                        & Loop    & Loop   & Loop   & Function & Loop   & Function & Function  \\
           Syscall arguments checked?             & \xmark  & \cmark & \xmark & \xmark   & \xmark & \xmark   & \xmark  \\
           Manually created BPF filter?              & \xmark  & \xmark & \cmark & \cmark   & \xmark & \xmark   & \xmark \\
           Lines of code added             & 19 & 43 & 298 & 25  & 57 & 58 & 45  \\
           Times documentation referred to & 12 & 2  & 18  & 8   & 8  & 4  & 10  \\
           Attempts running the code       & 9  & 8  & 19  & - & 20 & 14 & 30  \\
           \bottomrule 
       \end{tabularx}
     \label{tab:Seccomp_implementation}
 \end{table*}

 \subsubsection{Security awareness}

 All of our participants showed an interest in securing the program and tried to have a solution that would limit what attackers can do, and improve the security of the program. All of them showed motives toward achieving the \emph{principle of the least privilege}~\cite{Saltzer_Schroeder_1975} by stating that they wanted the program to have only the syscalls it \emph{needs} and nothing more. Despite all of them sharing the same motive and interest, their solutions and approaches were different. They all wanted the watchdog script to have only the necessary permissions, but the idea of \emph{what is necessary} was different across each participant.

 \begin{quote}
     ``I'm trying to restrict the process to only so that it can only use the syscalls that it needs.''\credit{P1}

     ``I want to prevent as much as possible.''\credit{P2}

 \end{quote}

 \subsubsection{What is necessary?}

 Seccomp is, ultimately, about choosing what syscalls to allow or deny. A simple program like the watchdog script will make use of many syscalls: some for setting up the program's memory and resources and some needed for the actual functionality of the program. The watchdog script functionality did not change dynamically, was not technically complicated, and did not use non-standard libraries or syscalls. Despite this, the set of syscalls the participants decided to operate on was different in all of the sessions. All of our participants who had a working solution, took an allow-listing approach\footnote{Sometimes called a \emph{whitelisting} approach.}, blocking everything by default and allowing only explicit syscalls, their motives behind this choice were that it is more secure, easier, and better for maintainability.

 \begin{quote}
     ``\ldots{}the disallow list doesn't work, It could work today but it won't work tomorrow.''\credit{P3}


     ``It's a lot easier to write a list of everything that is allowed than the entire universe of things that aren't allowed.''\credit{P6}
 \end{quote}

 All of our participants with a working solution allowed essential syscalls for the main loop of the  program to function, \texttt{openat}, \texttt{write}, \texttt{close}, \texttt{connect}, \texttt{socket}, \texttt{rename}, \texttt{clock\_nanosleep}, \texttt{newfstatat}, and \texttt{lseek}. P1 and P5 missed the \texttt{exit\_group} and did not allow it. The program needs this syscall in its exit path, when things go wrong, the program exits with an error message, however, these participants did not pay attention to the exit paths and only focused on the main program functionality. P3's main focus was on making the program portable by considering variations of syscalls across different architectures. For instance, besides allowing \texttt{newfstatat}, the participant also decided to allow \texttt{fstat} and \texttt{fstat64} syscalls, as different implementations of \texttt{libc} might use these syscalls when the \texttt{stat} function is triggered. The participant attempted to cover all existing syscalls variations by checking the operating system source code files and the man pages search. They allowed variations of \texttt{open}, \texttt{rename} and \texttt{write}, and also decided to account for systems that might use the \texttt{time} and \texttt{gettimeofday} syscalls instead of \texttt{clock\_nanosleep} and allowed them \emph{just in case} they were needed. 

 \begin{quote}
     ``It's a good thing to also allow \texttt{time} here, just to make sure.''\credit{P3}
 \end{quote}

 P5 and P7's sandboxes were over-privileged, as they allowed all syscalls that the program runs in its setup phase to allocate memory and setup the binary resources. P5 simply ran the program under strace---a syscall tracer for Linux that allows developers to inspect the syscalls a program makes---copied strace's output, and asked ChatGPT to write a Seccomp filter for it. P7 also allowed every syscall produced by strace but wrote the Seccomp filters by hand. These participants trusted strace's output and considered the outputted syscalls to be essential for the program. Their sandboxes allowed 15 syscalls that were not needed by the watchdog program, such as \texttt{ioctl}, \texttt{mmap}, \texttt{access}, and even \texttt{execv}. If an attacker managed to exploit their program, they would be able to run new processes, communicate over the network, read files, write files, and allocate memory, which makes their sandbox somewhat ineffective.

 \subsubsection{Where to install the sandbox?}

 Three of our participants decided to sandbox the whole program by installing Seccomp filters at the beginning of the main function, while four participants installed the Seccomp filters at a later stage to only sandbox the main loop. The only difference is that before the main loop, the program registers signal handlers using the \texttt{signal} \texttt{libc} function, which results in the \texttt{rt\_sigaction} syscall being triggered. When we asked participants about their decision of where they installed the sandbox, their answers varied.

 \begin{quote}
     ``It can be shifted to the start of the program to make it more secure, but I'm just concerned about this continuous tasks that are being performed. Signals aren't really hijackable. They're just registering a single event.''\credit{P1}


     ``In this simple case, we're using signal here, but we're not dealing with signals in the main. The the spot right before the the the main loop is usually the best place to turn down things''\credit{P3}

     ``I guess my thought was that I kind of wanted to do it before the program does anything else.''\credit{P6}
 \end{quote}

 \subsubsection{Good enough security}

 The sandboxes implemented by the participants did not just vary in terms of which syscalls they allowed, but also in whether they decided to examine syscall arguments. P1, P2, P3 and P6 were aware that Seccomp can examine syscall arguments. P5 said they knew about it but do not have enough experience with it.

 P2 was the only participant who decided to further restrict the sandbox and examine syscall arguments for the \texttt{write} and the \texttt{socket} syscall. The participant allowed the \texttt{write} syscall to operate on 3 file descriptors, the \texttt{stdout}, and the descriptors for the two log files. They also restricted the length of the buffer used by \texttt{write} syscall to only 32 bits or less. Their reasoning was that usually, buffer overflow vulnerabilities need to write long buffers, and the watchdog script only writes very short buffers, so limiting the length of the buffers adds a layer of security. They also restricted the \texttt{connect} syscall to only operate with \texttt{AF\_INET}, \texttt{SOCK\_STREAM}, and \texttt{IPPROTO\_IP} arguments, which means that the \texttt{connect} syscall cannot be used to connect to local \texttt{unix} sockets or use other protocols. 
 When we asked why they limited the arguments for \emph{these specific} syscalls, P2 responded that they were \emph{more dangerous} than others. Other participants considered some syscalls to be dangerous but did not agree on which. Some participants thought \texttt{rename} was dangerous as attackers could corrupt the file system with it. Other participants thought that \texttt{openat} syscall would be one they would like to limit and only allow it to operate on certain paths of the file system.

 \begin{quote}
     ``Maybe I want to compare the argument for the \texttt{signal} but it doesn't feel like something that necessarily is very important from a security perspective.''\credit{P6}



     ``As long the program can’t execute something, this shouldn't be a concern.''\credit{P1}    
 \end{quote}

 When we asked other participants why they did not limit the syscalls arguments their reasons, again, varied. Some thought it unnecessary, as many of these syscalls were \emph{harmless} under the current settings. Others stated that it was not worth the effort, as the program functionality is obvious, and maybe in a more critical scenario, they would consider it. P1 said that their sandbox provides adequate security and that as long as the program cannot execute, then it is secure. After finishing the task, P7 said that the current sandbox is \emph{good enough}, but they did not feel confident.

 \begin{quote}
     ``You can’t do much with \texttt{open} itself, I guess I can restrict it, but I think the allow filter is enough.''\credit{P1}

     ``We have a lot of things that are already blocked by default, and in this case restricting socket only to \texttt{AF\_INET}, is not doing very much.''\credit{P3}

     ``I personally don't feel confident about this implementation, but I think its good enough.''\credit{P7}
 \end{quote}

 Many of the participants considered the readability and maintainability of the program over security and having a restrictive sandbox. When asked why they did not restrict syscall arguments, they said that they would do this if they were sure the program would not change at all in the future, as any changes could lead to breaking syscall arguments filters, and that they wanted their sandbox to \emph{adapt less}.

 \begin{quote}
     ``If I only allow certain system calls, regardless of arguments, I  would have to adapt, much less.''\credit{P2}

     ``I probably wouldn't do it, because if we moved the server, and now it connects over IPV6, we have to keep changing the thing. We could do, but it's trade off, it doesn't buy us much.''\credit{P3}

 \end{quote}

 When we asked participants whether they considered restructuring the program so it opens resources (files and sockets) before the main loop, as this would enable installing a more restrictive sandbox, P1 said that the code was fine and there was no need to change anything, P2 said they did not do that because the time was tight, however, they had around 30 minutes remaining and they opted to tidy the code over restructuring the program as it would be complicated. P3 said that this is a good idea, however they did not consider it during the task, while P2 said that they would not want to change the program for Seccomp. 

 \begin{quote}
     ``I don't want to modify my entire program, just to be able to use Seccomp properly. If I want to use maximum security for a program, I will try to program it with that in mind.''\credit{P2}
 \end{quote}

 \subsection{Efficiency}

 Just as the developers implementations varied technically, the way they arrived at these implementations also differed, with different approaches, trains of thought and different resources.

 \begin{figure*}
     \centering
     \includegraphics[width=\linewidth]{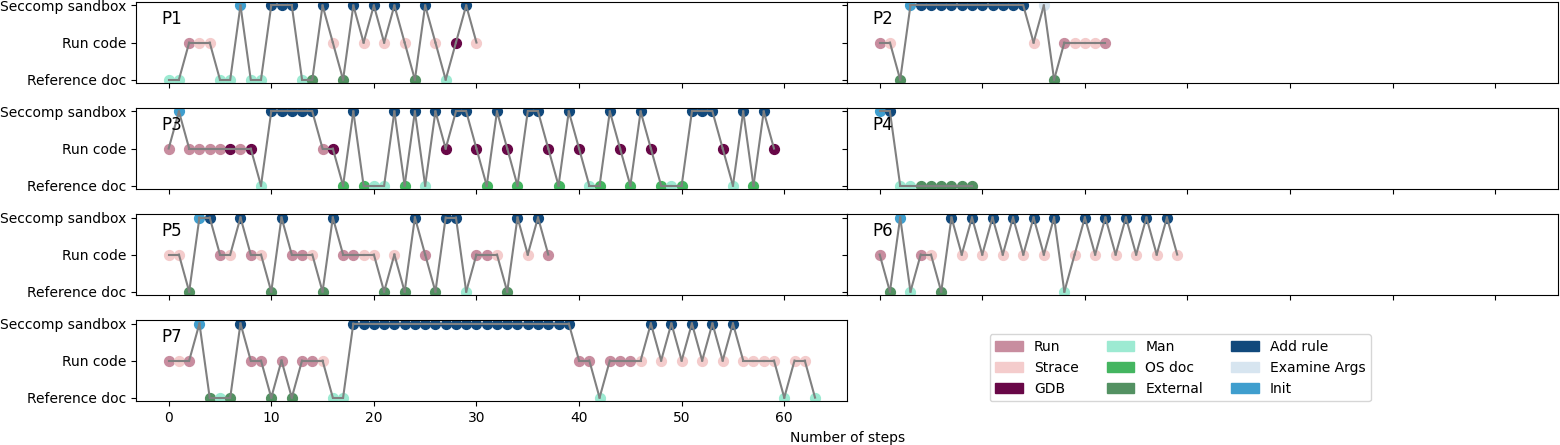}
     \caption{Timeline of participants steps when implementing the sandbox}
     \label{fig:Seccomp_approaches}
 \end{figure*}

 \subsubsection{A progressive approach}

 All bar P4 adopted a progressive approach to the task. They all started by reading the program, running it under strace, and then progressively adding Seccomp filters, checking if it worked, and repeating the process until they reached a state where they were happy with their solution, for some this process included frequently referencing documentation. 

 When adding Seccomp filters, some participants blindly trusted strace output and added all outputted syscalls, some even added syscalls that are needed only in the setup phase of the program and not needed afterward, for example, both P5 and P7 added a filter for \texttt{execv} and \texttt{mmap} syscalls, as strace outputted them, and they considered them to be needed. P7 tried to map strace output to the program functionality but they considered all the syscalls to be necessary.

 \begin{quote}
     ``All of these seems necessary to me\ldots{}all these \texttt{libc} wrappers make underlying system calls I don't even know about sometimes''\credit{P7}
 \end{quote}

 P2 adopted a different approach: they started by adding Seccomp filters based on a read of the source code, then they refined their list based on strace output and added other syscalls. P1, P3, and P6 did not add syscalls blindly and all mapped strace output to the program functionality before adding the filter. Sometimes they were not quite sure about where the syscall is used, and made guesses as to why it would be needed. For example, when P1 added a filter that allows \texttt{lseek} syscall, we asked them if they knew where is it being used in the application, they said they were unsure and ran the program in GDB just to check where is it being used, while P6 made a guess that it is used by \texttt{printf}. Later, both P1 and P6 figured out that it is used by \texttt{openat} because it opens the file in append mode.

 \begin{quote}
     ``I'm guessing that they are part of \texttt{fprintf}, but I am not sure.''\credit{P6}

     ``I don't have an obvious clue.''\credit{P1}
 \end{quote}

 Figure \ref{fig:Seccomp_approaches} shows the timeline of the steps each participant took to implement the sandbox.  Some participants, like P2, had a confident and smooth approach: they ran the program under strace once, added the needed syscalls, and then ran it again to check. Others, like P3 and P6, were more cautious adding some syscalls, running the program until it gave a \emph{bad syscall} error, and then repeat the process. P3 decided to search for variations of syscalls so they made frequent documentation lookups. P1 frequently looked up the man pages of \texttt{libc} functions and syscalls just to double-check their functionality. P7 trusted strace output and added all its syscall, and then later figured out that there are other syscalls missing and added them manually. P5 repeatedly asked ChatGPT to generate the filters and to add more syscalls to it.

 \subsubsection{Assuming mistakes}

 While solving the tasks, participants assumed that their sandbox implementation had mistakes from edge cases. P3 said that they would like to run their code on different architectures to make sure they covered all syscall variants, while P7 stressed the point that their sandbox needed checking as they might be exploited in a way that they do not know about. While writing the code, P2 grouped syscalls under categories such as networking and file-system-related syscalls, when asked why, they said that they assumed their sandbox would contain mistakes and this will make it easier for them to debug it later. When implementing the sandbox, P1 and P6 would anticipate that their sandbox would not work before running it, though were not able to guess which syscall would be denied. 

 The developers seemed to prefer to work in a progressive way and add a few rules at a time, run their program, and refine it later. They expected that their sandbox would not work until it \emph{actually} worked.  Whether this approach would work for a larger program is uncertain. P3 decided to keep the program running while doing the other phases of the interview, since, despite the simplicity of the program, and accounting for all the variations of syscalls, they \emph{still} were not sure if all syscalls were covered.

 \begin{quote}
     ``I think it's gonna give me a bad system call.''\credit{P1}

     ``I assume that I will make mistakes when implementing my Seccomp filter, so I want to make it easy for me to also fix it later and debug it.''\credit{P2}

     ``My assumption was that this would not work, but its working''\credit{P7}


 \end{quote}

 P1, P2, and P7 decided to double-check if their sandbox was working by making the program do things that should not be allowed under the sandbox. P1 decided to issue an \texttt{execv} syscall, while P2 tested the argument filter for the \texttt{write} syscall by making it write a large buffer; P2's test initially failed as they forgot to call the sandbox in the main function, but they subsequently found the bug and fixed it.

 \begin{quote}
     ``Let's make sure it actually does crash, because if it does not, then maybe something went wrong.''\credit{P2}


     ``Just to check the sanity of it, lets do something beyond these system calls.''\credit{P7}
 \end{quote}

 \subsubsection{Driving by example}

 Developers referenced man pages, blogs, and GitHub projects when working on the code. P3 came to the task prepared with a $\sim$200-line code wrapper for Seccomp; including macros and functions to use and debug Seccomp---as they wanted to reduce the number of dependencies their code needed. The participant's focus was on implementing a portable solution for multiple architectures and made heavy references to man pages and OS documentation to figure out variations of syscalls. P2, P6, and P7 started their implementations by copying code from blogs, modifying that code to fit their sandbox. P1 made references to the man pages to check what each library and syscall is doing before adding the filter. 

 P4 could not complete the task successfully. They started implementing their sandbox based on a previous implementation, copying an initial filter, and spent the rest of the time trying to figure out what exactly this filter was doing. As it used raw BPF filters, the participant was confused about some of its instructions. When becoming confused and frustrated, the participant just kept looking at different resources online, ranging from man pages, blogs, and GitHub source code, but could not satisfy their curiosity. After a while, they said that they did not think they could make any progress with the task, and that usually they like to take their time.

 \begin{quote}
      ``I suppose the way that I normally do things, I work example based, So I'll be looking up some of the Seccomp examples.''\credit{P4}



      ``So I'm gonna keep looking through some different maybe some easier Seccomp examples that show how Seccomp is being used''\credit{P4}


 \end{quote}

 P5 decided to rely on ChatGPT to implement his sandbox, he simply copied strace output and asked ChatGPT to create a Seccomp filter for him. The first time, ChatGPT produced a Seccomp filter that uses \texttt{SMPT\_ACT\_ALLOW} and also allows system calls from strace, which does not make any sense, since \texttt{SMPT\_ACT\_ALLOW} allows everything by default. The participant quickly noticed this and changed it to \texttt{SMPT\_ACT\_KILL}. But even with this, the program was not functioning correctly, and it would break after a few iterations because it issues a \texttt{rename} system call to rotate the log after it reaches a specific size, and as the first strace run does not capture this, ChatGPT solution does not allow it. After a while, we gave the participant a hint that the program could be doing things differently from the initial stage, and the participant caught the \texttt{rename} case immediately. 

 \vspace{\baselineskip}\noindent
 When we asked the participant why he decided to use ChatGPT, he said it would make things way faster.
 Ironically enough, P5 took the longest time of all the participants with a working solution.

 \subsection{Satisfaction}

 The participants expressed their struggles and frustration about various aspects of Seccomp. We captured these frustrations and used participants' answers to the SUS to reveal insights into how the developers feel about Seccomp's usability in general.

\subsubsection{A lot of stuff to track}

P3 came prepared with a library of macro definitions and helper functions that they used whenever working with Seccomp, as they felt that Seccomp could be quite verbose and some library code would save time. P1 and P2 both expressed that the Seccomp rules they had to add are long and hard to keep track of. 

\begin{quote}
    ``I prepared something because otherwise I would waste most of the time, just typing this stuff.''\credit{P3}

    ``These are a lot of Seccomp rules.''\credit{P1} 

    ``There's a lot of different calls and possible configurations that I have to keep track of.''\credit{P2}
\end{quote}

\subsubsection{Syscalls confusion}

One of the biggest issues participants faced was confusion about what syscalls the application uses and where. While strace outputs the list of syscalls that are triggered by the application, the participants were confused sometimes as they could not map where these syscalls are used in the application. P1 decided to launch the program in GDB to figure out if the \texttt{lseek} syscall is needed. P3 was debugging the program in GDB and noticed that it issues a \texttt{time} syscall. Later they figured out that this is a \texttt{VDSO} syscall and there was no need to filter it.

\begin{quote}
    ``Where is the \texttt{lseek} used?''\credit{P1}

    ``I don't see where we allow the the system call \texttt{time}, because \texttt{clock\_nanosleep} is to sleep.''\credit{P3}
\end{quote}

Sometimes developers were confused about the existence of syscalls; either because they had not seen them before, or because they could not figure out if they were library calls or syscalls. They checked the man pages and operating systems lists to make sure that these syscalls exist.

\begin{quote}
  ``I don’t think I have seen rename before.''\credit{P1}
  
  ``I just want to confirm that the \texttt{connect} system call actually exists.''\credit{P1}
  
  
  
  ``I have never heard of \texttt{rename} before, I guess its new to me.''\credit{P7}
  
  ``It's not usually that hard, but when you mix all the libraries, external libraries, and all libadmin, or some fancy old libraries could get more more difficult.''\credit{P3}
  
  ``\texttt{newfstatat}, that's one I didn't expect.''\credit{P6}
\end{quote}

After implementing the sandbox, P7 carried out additional tests and noticed that the program was terminated by Seccomp as it tried to run \texttt{restart\_syscall} syscall. This is confusing as the core functionally of the program does not use this syscall and it is internally used by the kernel to restart certain syscalls to adjust their time-related operations (in this case, \texttt{clock\_nanosleep}). The participant was very confused as they could not figure out where this syscall is used and could not replicate the issue.

\begin{quote}
    ``I am not hitting the \texttt{restart\_syscall} condition again, I don't know why that happened\ldots{}''\credit{P7}
\end{quote}

\subsubsection{Setup syscalls}

Participants expressed that sometimes it is difficult to distinguish between the setup syscalls and the ones the application needs. As discussed before, P5 and P7 implemented a sandbox that allows for all the syscalls strace outputs, resulting in an over-privileged sandbox that allows for all the setup syscalls. P1 decided to compile the program statically, to eliminate these setup syscalls and reduce the noise printed by strace. 

\begin{quote}
    ``It's going to give me these useless library loading systems, and I have to search strace logs.''\credit{P1}
    
    ``There might be some system calls that are only used during setup. I have to make sure I don't disallow the normal function.''\credit{P2}
\end{quote}

\subsubsection{There should be a better way to do this}

Many of the participants felt there should be a better way of doing this, even though they were not doing anything wrong at the time. P2 felt that manually examining syscall arguments is not optimal and should be automated when done with larger programs. Whilst copying and pasting filters for syscall variants, P3 said that there must be a better way of doing this. P6 said that there had to be another way to attach different Seccomp filters to each application component.

\begin{quote}
    ``Of course, with larger programs doing this manually is something that I wouldn't want to do, because that's a lot of effort. I think this will already be quite difficult to protect this relatively small program.''\credit{P2}

    ``Maybe there are other ways to do this.''\credit{P3}
\end{quote}

\subsection{SUS results}
\begin{figure*}
  \centering
  \includegraphics[width=.95\linewidth]{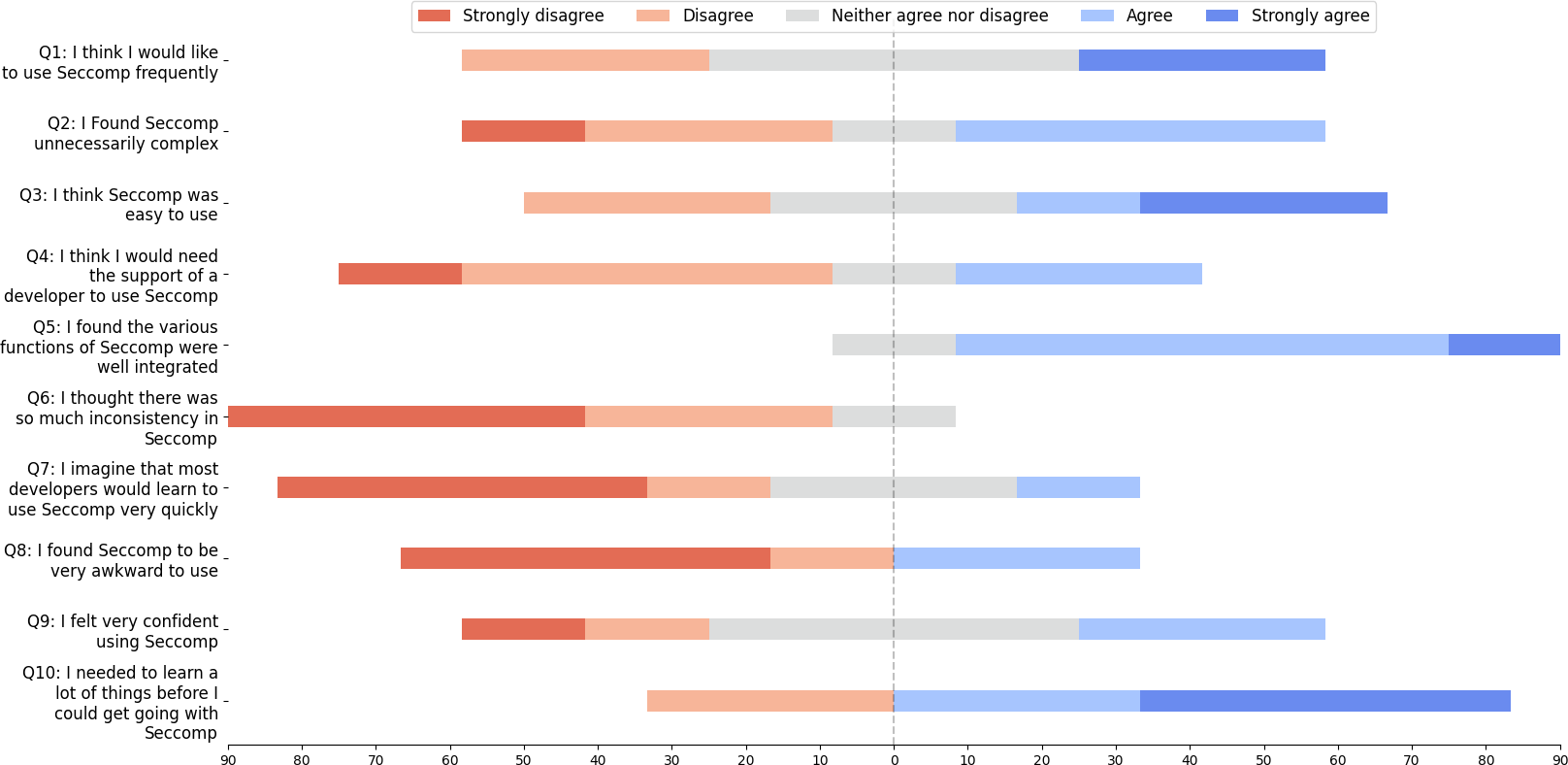}
  \caption{Answers to SUS questions.}
  \label{fig:sus_answers}
\end{figure*}

The participants' responses to the SUS are shown in Figure \ref{fig:sus_answers}. We asked the participants to expand on their answers verbally. The participants had different opinions on whether they would use Seccomp frequently or no, some said that they would, and some said that their work does not require it.

\begin{quote}
    ``I think I would like to use Seccomp frequently, it can be very easy to use.''\credit{P1}


    ``It is a bit of a pain to use and to debug, and there are all sorts of differences in architectures and existing causes.''\credit{P3}

\end{quote}

When asked if Seccomp is unnecessarily complex, and if it is easy to use or not, participants also had different thoughts. Some said that once you understand Seccomp, it is easy to use, but it would be hard to approach without prior knowledge. Others said that they understand why Seccomp has its complex interface, and that it is not \emph{unnecessarily} complex. All bar P3 said that they did not need the support of another developer while using Seccomp, P3 instead thought that  the support of another developer would definitely help in figuring out the needed syscalls across architectures and also remind them of steps like checking arguments.

\begin{quote}
    ``Once you understand it, it's fine, but approaching it without free knowledge, it's not easy.''\credit{P2}


    ``I don't believe was easy to use. It requires the knowledge of the underlying implementation of the C library.''\credit{P3}
\end{quote}

Participants agreed that the various features of Seccomp are well integrated and that Seccomp is consistent. P6 thought that the API made sense, as all you have to do is initialize the filter and add rules. P2 and P3 said that Seccomp does not have many functions to begin with and the ones they used in the task are well integrated. 

\begin{quote}
    ``It is hard to use. It is overly complicated, I believe, but it isn't inconsistent.''\credit{P3}
\end{quote}

In their answers to Question 7 and Question 10, most of the participants thought that developers need to learn a lot of things to use Seccomp, as they also themselves needed to learn a lot of things before using Seccomp, mainly because it needs an understanding of the concept of syscalls, and of system internals. 

\begin{quote}
    ``If we really just talk about developers and various backgrounds. I would imagine that it's very, very difficult.''\credit{P2}

    ``You have to learn the byte code. You have to learn how this library works underneath.''\credit{P3}

     ``I was able to sort of get going with it fairly easily but as far as like integrating into a real program, I imagine
     it would be more complicated.''\credit{P6}
\end{quote}

When asked if they felt confident while using Seccomp, the answer varied: several expressed a lack of confidence.

\begin{quote}
    ``I don't feel confident in terms of understanding what it does or how it does it.''\credit{P4}

    ``I do feel like I understand how Seccomp works in general. I don't think I know all of the edge cases and all of the features it has to offer.''\credit{P2}

     ``It depends on what you mean with confident, if we mean that the program won't do anything except what I told it to do. It's a strong 5. Let's say a 3 only because there are very interesting differences between the architectures''\credit{P3}
\end{quote}

\subsection{Design}

In the last stage of the interview, we asked participants to describe a sandboxing model they would use always, one that fits their needs, we asked them to design the \emph{perfect} sandbox. 

\subsubsection{Abstraction layer}

Some participants thought that the Seccomp model of sandboxing is a good one and that they would stick to a model that is based on the syscall interface, but others designed a sandbox that would have an abstraction layer that hides the complexity of syscalls.

\begin{quote}
    ``System calls are a very convenient way of interacting with the kernel.''\credit{P1}


    ``It's simpler to think about in in some slightly more high level groups. it's difficult for a developer to think in terms of code using \texttt{newfstatat}, \texttt{fstat64}\ldots{}etc.''\credit{P3}
\end{quote}

P4 was motivated by the need for a sandboxing mechanism that helps in their embedded-systems work where sometimes they need to execute untrusted user scripts. Their proposal is shown in Listing \ref{lst:P4 design}. They opted for an abstraction layer where developers can assign permissions such as socket communication and file access to certain scripts before they run.

\code{P4 design}{XML}{code/p4_design.py}

\subsubsection{Automation}

P2 thought that automation was needed for Seccomp to be usable by the average developer. When we asked the participant if they were aware of any tools that automate Seccomp filtering, they said they were not, but in fact, there is a lot of research around this area and many tools have been developed in both the academic research and by the open source community to automate Seccomp filtering~\cite{canella2021automating, demarinis2020sysfilter, gelderie2022seccomp}. With that being said, there is no data regarding evaluating the adoption and accuracy of such tools.

\begin{quote}
    ``I mean, if we want to implement something that can be used by average developer, then there has to be some level of automation or as some tools that can assist here.''\credit{P2}

    ``If the developer has some way of some semi automated way of acquiring, of getting all of the [syscalls] that are involved in the no-program execution\ldots{}''\credit{P2}
\end{quote}

\subsubsection{Block dangerous code}

Many participants suggested a sandboxing mechanism that is focused on containing dangerous areas, and many ended up designing a container model where they would contain \emph{dangerous} code. The container model came up many times and it seems to be appealing to the developers.

\begin{quote}
    ``You can't run \texttt{execv}. You can't run the dangerous system calls.''\credit{P1}


    ``I would definitely focus on dangerous calls.''\credit{P2}


\end{quote}

\subsubsection{Sandboxing logic closer to the code}

While solving the task, after getting a working solution with Seccomp, P6 started refining their code, and wrote the function \texttt{allow\_write\_log} to contain the Seccomp rules for the \texttt{write\_log} function. The idea was that instead of having a long function that contains all the Seccomp rules for the whole application, instead, P6 attempted to split the Seccomp rules list into different functions, each function would contain the filters for a specific function in the code. They \emph{annotated} the \texttt{write\_log} function with the syscalls it needed (Listing \ref{lst:P6 design}).

In the design phase, the participant expanded on their approach by stressing the point that they would like to have \emph{more locality} between the code being sandboxed and the sandbox policy and wondered if there is a way to have different filters, one for each component of the application.

\code{P6 design}{C}{code/p6_design.c}

\begin{quote}
    ``It would be nice to be able to put some of the the like sandboxing logic closer to the code.''\credit{P6}


    ``I'm not really sure how to draw this, but more locality between the the filtering and the code being filtered versus having all of the filtering in one location.''\credit{P6}

\end{quote}

Similar points were made by P4 since they suggested a wrapper approach, where selected code can be wrapped with \texttt{sys\_blocker}, the sandbox designed by the participant is shown in Listing \ref{lst:P4 design}. When asked about which components would be wrapped inside \texttt{sys\_blocker}, they said that it would be attached to functions.

\begin{quote}
    ``I think if I was to implement something like this, I would probably implement something that would act more as a wrapper.''\credit{P4}

    ``I would imagine it wrapping functions. I think maybe at the end of the day functions are the things that are doing these actions, opening or closing, or forking.''\credit{P4}
\end{quote}

\section{Discussion}

Security and usability is a trade-off. A hard to use sandboxing mechanism would not be adopted by developers; but Seccomp seems flexible to a fault. You can implement allow or deny lists, configure it with the syscalls you \emph{think} are needed, add variations for other architectures, and examine arguments. Custom BPF filters can create a fine-grained sandbox for an application, but this power leads to many implementation choices the sandbox and the possibility that things can go wrong.  It places more responsibility on the shoulders of developers to make these decisions.

Does that make Seccomp a bad sandboxing mechanism? It depends on whom Seccomp is intended to be used by, and why it was created in the first place. In the kernel mailing list, Chromium developers discussed Seccomp's filter mode to benefit Chrome sandbox. Their idea was opposed as a maintenance nightmare~\cite{Kernellist2011}.

\begin{quote}
    ``Security is a morass. People come up with cool ideas every day, and nobody actually uses them---or if they use them, they are just a maintenance nightmare\ldots{}And per-system-call permissions are very dubious\ldots{}What system calls don't you want to succeed? That \texttt{ioctl}? You just made it impossible to do a modern graphical application.''\credit{Linus Torvalds}
\end{quote}

Seccomp relies on the system call interface, which is inherently complex, varies across systems, and requires developers to reason about low-level details. However, some usability issues also stem from Seccomp’s current implementation, for example, a lack of helper tools or automation to guide developers through common use cases and best practices.
While Seccomp provides rich functionality to implement sandboxes, the complexity of the syscall interface and the many decisions involved in using it can result in many usability issues. Pledge from OpenBSD takes a completely different approach to sandboxing, as it has an abstraction layer and can be integrated into applications with one line of code. While Pledge offers less fine granularity than Seccomp, it could result in a much better user experience.

\subsection{Sandboxing functions}

Some developers had a function-driven mindset---they would focus on the sandbox and security of each individual function in the application. P1 started by reading the program function by function, and sometimes they would state that some functions are \emph{safe}. The idea of sandboxing specific functions has not been discussed much, but such a model might fit developers' mental-models better. It would allow them to sandbox specific modules and focus on figuring out the needed permissions for each function, rather than the whole application at once. Combining this with an abstraction layer that hides the low-level details of the sandbox may also improve usability and make sandboxing programs less burdensome.

\subsection{The system call interface}

Is the problem that Seccomp uses the system call interface to impose the sandbox policy? The participants did not agree on the same set of system calls to restrict, some allowed unnecessary system calls, and others accounted for many that could be used under different systems. The usability challenges also seem to mainly stem from the variety and complexity of the system call interface, as many of the developers could not map the functionality of these system calls to the program being sandboxed leading them to be confused and less confident about their implementations. 

In his research, Garfinkel et al. highlighted many technical challenges that arise when interposing on the system call interface and questioned whether it is a good place to implement the sandbox policy \cite{garfinkel2003traps}. Other solutions like Pledge from OpenBSD provides a different approach to sandboxing, as it has an abstraction layer that is easy to use and does not require inspecting every system call an application makes.

\subsection{Sandboxing design patterns}

In  software engineering design patterns and standards (such as the SOLID principles) can guide developers into implementing  robust code and provide a framework that can be used to determine good implementations from bad ones~\cite{plosch2016measuring}. `
The same cannot be said of sandboxing: it is not always clear how you should implement them. 

Whilst Seccomp focuses on filtering system calls, Pledge simplifies restrictions through a high-level interface, while Capsicum (from FreeBSD) emphasizes capability-based security. The lack of a unified approach means that developers must navigate a fragmented landscape, re-implementing similar functionality with different tools. Even when considering just Seccomp, there is more than one way to build a sandbox and the design trade-offs are not obvious (even to experienced developers).

The lack of standard design patterns leads to practical challenges. It can lead to inconsistent security practices, where the effectiveness of sandboxing implementation is highly dependent on the specific API and the developer's familiarity with it harming the overall reliability of sandboxing as a security strategy. The integration of sandboxing into existing systems can become cumbersome and error-prone, with each API requiring different integration techniques and considerations. Could there be a framework or design principles for sandboxing that promotes consistency, ease of use, and robust security? How can we develop a more unified approach to sandboxing that leverages the strengths of existing mechanisms while addressing their individual shortcomings?

Even when sandboxing simple programs, developers end up with different sandboxes.  Our study establishes a starting point for building design patterns for sandboxing: the progressive approach taken by several developers could be abstracted into a pattern,  allowing for modular sandboxing.

\subsection{Artificial \emph{intelligence}}

When ChatGPT was released some of its users were able to get it to generate malicious code and exploits. Later it was tweaked to not generate such code. P5 used ChatGPT to generate Seccomp policies to sandbox the application: ChatGPT generated a policy that allowed the \texttt{execv} syscall, the developer trusted the generated policy, although it was flawed and over-privileged.

For security code: is giving users flawed security policies as bad as generating an exploit? Writing flawed code that is perceived to be secure has catastrophic effects. Seccomp policies are not easy to verify, and can be long, contain low-level and system-specific details. These can be difficult to check for correctness. The overwhelming task of reading strace output and writing long lists of rules could be why P5 used ChatGPT instead of configuring the sandbox themselves.

\section{Conclusion}


Seccomp is a powerful tool for sandboxing, but we lack standard ways to apply it. Developers flounder around trying to sandbox their applications, and become confused by syscalls.  By providing standard methods to building sandboxes and fitting the abstractions to developers mental models about how their programs should work we can move from \emph{playing in the sandbox} to having sandboxes being a standard mechanism incorporated into all programs.

\footnotesize \bibliographystyle{plain}
\bibliography{bibliography}

\appendix
\section{Watchdog script code}
\label{appendix:code}
\code{Interview task code}{C}{code/task.c}

\newpage
\section{Study Protocol}
\label{appendix:protocol}
\subsection{Beginning of interview}
\begin{itemize}
\item Make sure that the participants signed the PIS and Consent forms.
\item Make sure that the participants can access the VM and have no issues.
\item ``Hello, thank you for participating in this study. I will start by describing the task and
  interview''.
\item ``In this interview, you will be tasked with sandboxing a C program that simulates a
  watchdog service using Seccomp. You will have 1-hour to complete the task, and then
  we will ask you to fill out a form and participate in a design exercise. We will ask you to
  share your screen while doing so, and the audio of the interview will be recorded. You
  will have SSH access to a remote virtual machine for completing the task. Do you have
  any questions?''
  ``Voucher will be sent to you in the next 24-hours after the interview is completed.''
\item ``Let’s now make sure you have a connection to the virtual machine, and let me know if
  you would like any editors to be installed or any programs. I believe I sent you a
  document with connection details, have you had the chance to check it out?''
\item Ask participants to share their screen and show access to VM.
\item Ask participants to think out loud while they are solving the task. ``I am interested in your
  thought and decision making process, and I would love to hear it. Please think out loud
  and speak your mind.''
\item ``We can begin the study now''
\end{itemize}

\subsection{Coding task (1h)}
\label{sec:coding-task-1h}

\begin{enumerate}
\item ``I will start the audio recording now''.
\item Read the task scenario
\item Take note of and keep a timeline of:
  \begin{itemize}
  \item Every time participants refer to the documentation.
  \item Every time participants modify the Seccomp sandbox (initialize the sandbox,
    change rules/examine arguments).
  \item Every time participants run the code and how they run it.
  \end{itemize}
\end{enumerate}

\subsubsection{Scenario}
\label{sec:scenario}

``You are the system administrator of a server infrastructure, and one of your responsibilities is to
ensure the security and stability of various background processes running on the server. One
such process is the Watchdog script, a critical maintenance tool that performs tasks such as log
rotation and recording data. The Watchdog script is written in C and runs as a background
process. It performs tasks at specific intervals, such as checking if a port is open in a remote
server, rotating logs, and handling signals. It runs indefinitely and performs its tasks at regular
intervals.

As the system administrator, you are increasingly concerned about the security of the server,
especially given the potential risks associated with that background process, since it
communicates with an external server, and that it is written in C, where the program can have
memory vulnerabilities that can lead to exploitation. Given these security concerns, you
recognize the need to implement additional security measures. Your task is to enhance the
security of the Watchdog script and sandbox it using Seccomp, addressing the identified
security concerns and implementing a more robust security posture for this critical background
process.

As you can see, the watchdog script is in your home directory, and there is also a Makefile to
compile it, feel free to modify this code in any way you would like to to implement the sandbox. I
am interested in hearing your thought process and suggestions or struggles. Please think out
loud.

Unless you have any questions, You can begin the task now.''

\subsubsection{When to prompt/ask participants}
\label{sec:when-prompt-part}

\begin{enumerate}
\item When the participants reach a solution they are happy with, or at the end of the task time
  ask about:
  \begin{itemize}
  \item Decision on allow/deny list approach. ``Why did you choose the allow/deny listing
    approach in your solution? What is your reasoning behind this decision?''
  \item Decision on examining syscall arguments. ``Why did/did not you use
    Seccomp to restrict syscalls’ arguments?''
  \item Decision on where they placed Seccomp filters. ``Can you expand more on the
    reasoning behind where you installed the Seccomp filters in the code?''
  \item Decision on restructuring the program to make the sandbox more restrictive.
    ``Did you consider restructuring the program code to enable a more restrictive
    sandbox?''
  \end{itemize}
  \item When seems to be stuck for a while
  \item When the participant exhibits an interesting behaviour, ask about the reasoning behind
  this behaviour and try to note their train of thoughts.
  \item When the participant refers to a documentation, ask about what they are looking for
  exactly and why? ``Can you tell me what are you looking for and why''?
\end{enumerate}

\subsection{Filling in the SUS form ($\sim$15 minutes)}
\label{sec:filling-sus-form}

\begin{enumerate}
 \item ``The next stage of the interview will be filling a form. I will send you the form link in the
  meeting chat. Please let me know if you can access the form. The form has 10
  statements, you can rank each statement with a number from 0-5, 0 means that you
  strongly disagree, while 5 means that you strongly agree with the statement. While you
  are filling the form, please explain your answer and expand on it more.''
\item Let the participants describe their answers fully and ask follow up questions if any
  interesting point came up or if answers seemed vague.
\end{enumerate}

\subsection{Design exercise ($\sim$15 min)}
\label{sec:design-exercise-}

\begin{enumerate}
\item ``The last stage of the interview is a design exercise. I will share a white board with you
  now. Can you check if you can draw on the white board and access it without any
  problems?''
\item ``I am going to ask you to design a sandboxing model that can be used within a
  codebase (API), a sandboxing model that you would use frequently, the perfect sandbox.
  The sandbox should be able to restrict what your code can do. It is up to you how to
  design the sandbox, you can describe it verbally or by drawing on the board. I am more
  interested in the general design and of the sandbox rather than its technical
  implementation.''
\item After the participants reach an initial design, ask about why they chose this model? ``Can
  you tell me why you chose this model of sandboxing?''
\item Ask participants clarifying questions if their design seems unclear. ``Can you expand
  more on \ldots{} please?'' ``What do you mean by\ldots?
\end{enumerate}

\subsection{End of the interview}
\begin{itemize}
 \item Thank the participant and start the process of issuing the voucher.
\item Make sure the audio recording is saved.
\item Make sure the written notes are dated and clear.
\item Save a copy of the participant code solution.
\item Make sure the SUS form is submitted.
\item Save the content of the whiteboard.
\item Terminate the SSH account on the VM.
\end{itemize}

\newpage
\section{Recruitment Material}
\label{appendix:recruitment}

The following message was used to recruit developers to the study through a variety of channels including forum posts and developer mailing lists.

\begin{quotation}
  \noindent\textbf{Title: C Developer with Seccomp experience}

  We are
looking for C developers who have experience with Seccomp for a 2
hours study where you will be asked to use Seccomp to sandbox a simple
C program. This is a fixed-price project \redact{(200£)}.

To be able to participate in this study, you have to:
\begin{itemize}
\item Have C programming experience.
\item Have experience with Seccomp.
\item Be able to use command-line editors or setup your own editor via
SSH.
\end{itemize} You will be given SSH access to a virtual machine where
you will be solving the task, we will ask you to share the screen and
the audio of the session will be recorded and later anonymised by
removing individual names, and identifying information. The study will
take 2 hours maximum, and the job will be completed once the session
is over.

To apply for this study, please answer the following questions:
\begin{enumerate}
\item Tell us more about your C programming experience, what kind of
projects have you worked on before that required C programming?
\item Have you used Seccomp before? What kind of projects and why you
used Seccomp there?
\end{enumerate} This study aims to evaluate the usability of Seccomp
and is part of \redact{a PhD project}
\end{quotation}

\end{document}